\documentclass[conference]{IEEEtran}
\IEEEoverridecommandlockouts
\usepackage{cite}
\usepackage{amsmath,amssymb,amsfonts}
\usepackage{algorithmic}
\usepackage{float}  
\usepackage{graphicx}
\usepackage{textcomp}
\usepackage{xcolor}
\usepackage{subfigure}
\usepackage{multirow} 
\usepackage{booktabs}
\usepackage{adjustbox}
\usepackage{authblk}
\def\BibTeX{{\rm B\kern-.05em{\sc i\kern-.025em b}\kern-.08em
    T\kern-.1667em\lower.7ex\hbox{E}\kern-.125emX}}
\usepackage{algorithmic}
\usepackage{algorithm}

\usepackage{hyperref}
\usepackage[capitalize]{cleveref}

\usepackage{xspace}
\makeatletter
\DeclareRobustCommand\onedot{\futurelet\@let@token\@onedot}
\def\@onedot{\ifx\@let@token.\else.\null\fi\xspace}

\def\etal{\emph{et al}\onedot}
\makeatother

\title{
Elevating Medical Image Security: A Cryptographic Framework Integrating Hyperchaotic Map and GRU
}

\author[1]{Weixuan Li}
\author[2,3]{Guang Yu}
\author[1]{Quanjun Li}
\author[1]{Junhua Zhou}
\author[1]{Jiajun Chen}
\author[2]{Yihang Dong}
\author[2]{\authorcr Mengqian Wang}
\author[2*]{Zimeng Li\thanks{* Corresponding authors: li\_zimeng@szpu.edu.cn, xuhangc@hzu.edu.cn}}
\author[2]{Changwei Gong}
\author[4]{Lin Tang}
\author[5*]{Xuhang Chen\thanks{This work was supported in part by the National Natural Science Foundation of China (Grant No. 62501412), in part by Shenzhen Medical Research Fund (Grant No. A2503006), in part by Shenzhen Polytechnic University Research Fund (Grant No. 6025310023K) and in part by Guangdong Basic and Applied Basic Research Foundation (Grant No. 2024A1515140010).}}

\affil[1]{Guangdong University of Technology}
\affil[2]{School of Electronic and Communication Engineering, Shenzhen Polytechnic University}
\affil[3]{Guangzhou City University of Technology}
\affil[4]{Fuzhou University Zhicheng College}
\affil[5]{School of Computer Science and Engineering, Huizhou University}

\begin{document}
\maketitle

\begin{abstract}
Chaotic systems play a key role in modern image encryption due to their sensitivity to initial conditions, ergodicity, and complex dynamics. However, many existing chaos-based encryption methods suffer from vulnerabilities, such as inadequate permutation and diffusion, and suboptimal pseudorandom properties. This paper presents Kun-IE, a novel encryption framework designed to address these issues. The framework features two key contributions: the development of the 2D Sin-Cos Pi Hyperchaotic Map (2D-SCPHM), which offers a broader chaotic range and superior pseudorandom sequence generation, and the introduction of Kun-SCAN, a novel permutation strategy that significantly reduces pixel correlations, enhancing resistance to statistical attacks. Kun-IE is flexible and supports encryption for images of any size. Experimental results and security analyses demonstrate its robustness against various cryptanalytic attacks, making it a strong solution for secure image communication. The code is available at this \href{https://github.com/QuincyQAQ/Elevating-Medical-Image-Security-A-Cryptographic-Framework-Integrating-Hyperchaotic-Map-and-GRU}{link}.
\end{abstract}


\section{INTRODUCTION}
\label{sec1}
The rise of medical imaging and digital communication highlights the need for strong security in transmitting medical images, which often contain sensitive patient information vulnerable to unauthorized access and tampering. Traditional encryption methods, designed for text, are insufficient for medical imagery due to high inter-pixel correlation and large data volumes \cite{chen2019medical}. As a result, researchers are increasingly turning to chaotic systems to develop efficient and secure image encryption schemes.

Chaotic systems, characterized by their inherent unpredictability and acute sensitivity to initial conditions, exhibit properties that align closely with fundamental cryptographic requirements. This has led to their widespread adoption in generating pseudo-random sequences crucial for image encryption processes. While simple one-dimensional (1D) chaotic systems have gained popularity owing to their low computational complexity \cite{huang2019symmetric,le2024medical}---exemplified by Huang \etal \cite{huang2019symmetric} who proposed a permutation-diffusion simultaneous operation (PDSO) leveraging key streams from three 1D chaotic systems---they are often plagued by issues of predictability and vulnerability to cryptanalytic attacks \cite{xiaofu1999general}. Conversely, higher-dimensional chaotic systems, though offering enhanced security, typically impose computational burdens that render them impractical for resource-constrained platforms \cite{hua2016image}. Two-dimensional (2D) chaotic maps emerge as a compelling compromise, striking an effective balance between security and computational efficiency. Numerous such systems have been proposed for image encryption; for instance, Hu \etal \cite{hu2023novel} introduced the 2D-SSCDB map, which facilitates simultaneous scrambling and diffusion. More recently, deep learning has surfaced as a novel paradigm in this domain. Notable contributions include Google Brain’s self-encryption model employing adjoint networks \cite{he2021new}, LSTM-based methodologies for evaluating randomness \cite{he2021new}, and hybrid models integrating neural networks with chaotic systems \cite{zhao2021new}. Despite their potential, deep learning-based techniques are still in their nascent stages of development \cite{li2022few,li2025adaptive,liu2024dh,li2022monocular,liu2024depth,li2023cee,liu2023explicit,zhu2024test,li2024cross,liu2023coordfill,zheng2024smaformer}, particularly concerning color image encryption, which introduces complexities arising from inter-channel dependencies and significantly larger data volumes.

Effective pixel permutation is paramount in image encryption, serving to disrupt statistical regularities within images and thereby fortify defenses against histogram-based and correlation-based attacks \cite{xingyuan2019image}. Scanning-based permutation techniques, such as zigzag \cite{xingyuan2019image}, raster scan \cite{jolfaei2012image}, and spiral scan \cite{tang2019image}, are extensively employed. These methodologies linearize image pixels along predefined trajectories before reordering them to obscure discernible patterns. However, certain scanning approaches exhibit inherent weaknesses, including periodicity, incomplete pixel rearrangement, and susceptibility to cryptanalysis \cite{xiao2022image}. For instance, vulnerabilities in raster-scan-based encryption systems have been demonstrated by Cao \etal \cite{cao2020designing} and Shi \etal \cite{shi2022security}. To surmount these limitations, this paper introduces Knot-like Unique Novel-Scan Algorithm (Kun-SCAN), a dynamic scanning methodology that synergistically integrates chaotic nonlinearity with cross-region pixel reorganization. Kun-SCAN is designed to dismantle periodic patterns, augment randomness, and bolster resilience against statistical and brute-force attacks, thereby substantially enhancing the security and robustness of image encryption schemes.

The main contributions and unique aspects of this study are summarized as follows:
\begin{enumerate}
    \item \textbf{A novel 2D hyperchaotic map, termed 2D-SCPHM:} This map, leveraging sine and cosine functions with dynamic parameter adjustment, overcomes the limited chaotic ranges and few control parameters of conventional maps, generating sequences with enhanced randomness and wider chaotic ranges suitable for robust key generation.

    \item \textbf{A novel chaotic sequence generator employing the proposed 2D-SCPHM with Gated Recurrent Units (GRU):} This approach further explores the synergy between advanced chaotic maps and recurrent neural networks, demonstrating the versatility of 2D-SCPHM in generating highly secure key streams when coupled with GRU architectures\cite{chung2014empirical}.

    \item \textbf{A dynamic and robust pixel permutation algorithm, Kun-SCAN:} Addressing the vulnerabilities of traditional repetitive scan paths, Kun-SCAN divides the image into regions and implements chaotic-driven, cross-boundary pixel reorganization. This method effectively reduces local pixel correlations and significantly enhances resistance to statistical and brute-force attacks, thereby improving overall encryption security.
\end{enumerate}

\section{METHODOLOGY}
\label{Sec2}
This section introduces Kun-IE, a multi-image encryption algorithm with three core modules: (i) a chaotic sequence generator using the 2D Sin-Cos Pi Hyperchaotic Map (2D-SCPHM) and a Gated Recurrent Unit (GRU); (ii) the Knot-like Unique Novel-Scan Algorithm (Kun-SCAN) for pixel scrambling; and (iii) the XOR Diffusion Algorithm (XDA) for pixel value modification. Kun-IE enables concurrent encryption of multiple images, with security relying on secret keys $Key_{x0}$, $Key_{y0}$, $Key_{a}$, $Key_{b}$, and $Key_{N0}$. The following subsections detail each module.

\begin{figure}[ht]
    \centering
    \includegraphics[width=1\linewidth]{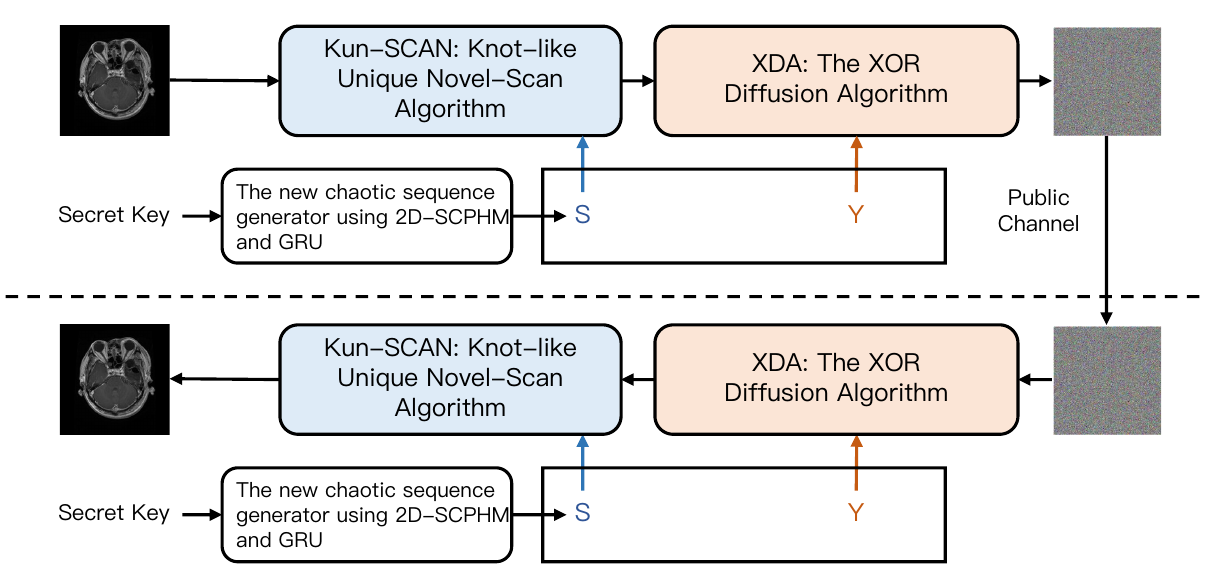}
    \caption{The process of the encryption algorithm.}
    \label{process}
\end{figure}

\subsection{Chaotic Sequence Generation using 2D-SCPHM and GRU}
\label{section2}

The proposed chaotic sequence generator combines a Gated Recurrent Unit (GRU) with a novel 2D Sin-Cos Pi Hyperchaotic Map (2D-SCPHM). GRUs, which model temporal dependencies using gating mechanisms, enhance chaos by amplifying small perturbations and driving divergence in phase space \cite{cho2014learning, hochreiter1997long}. The update gate’s memory management allows GRUs to capture complex, nonlinear dynamics, making them ideal for chaotic systems. Our approach leverages GRU's memory and nonlinear processing, alongside 2D-SCPHM, to generate sequences with superior randomness.

\begin{figure}[ht]
	\centering
	\includegraphics[width=\linewidth]{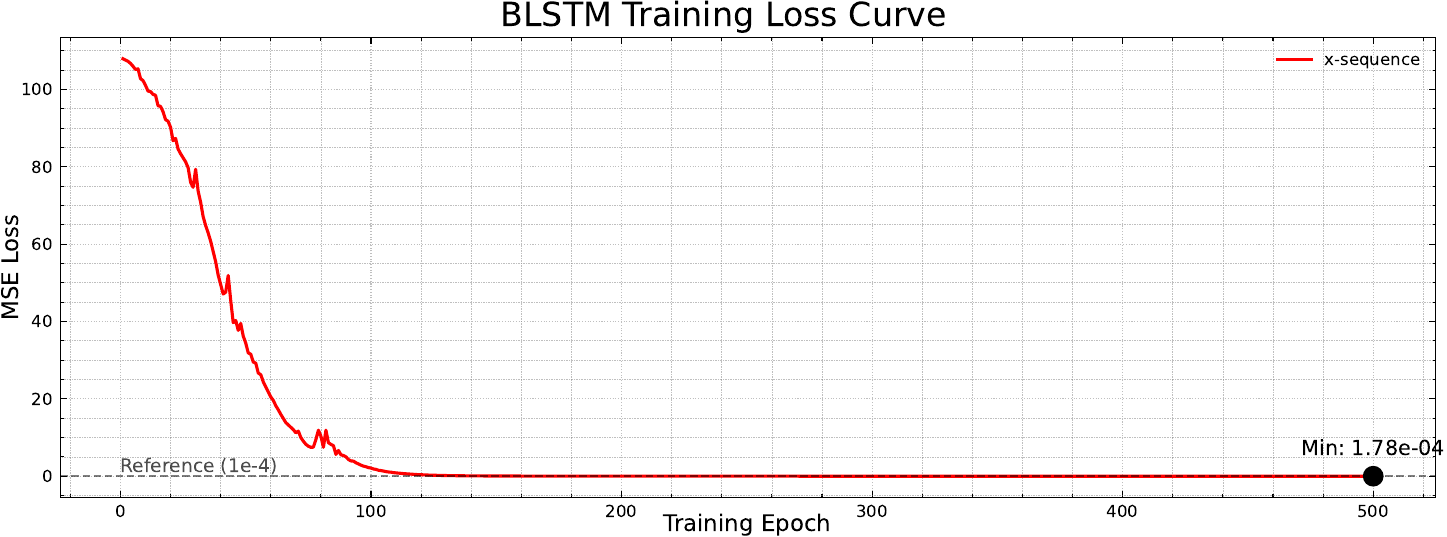}
	\caption{The fission diffusion}
	\label{The fission diffusion}
\end{figure}  

\begin{figure}[ht]
	\centering
	\includegraphics[width=\linewidth]{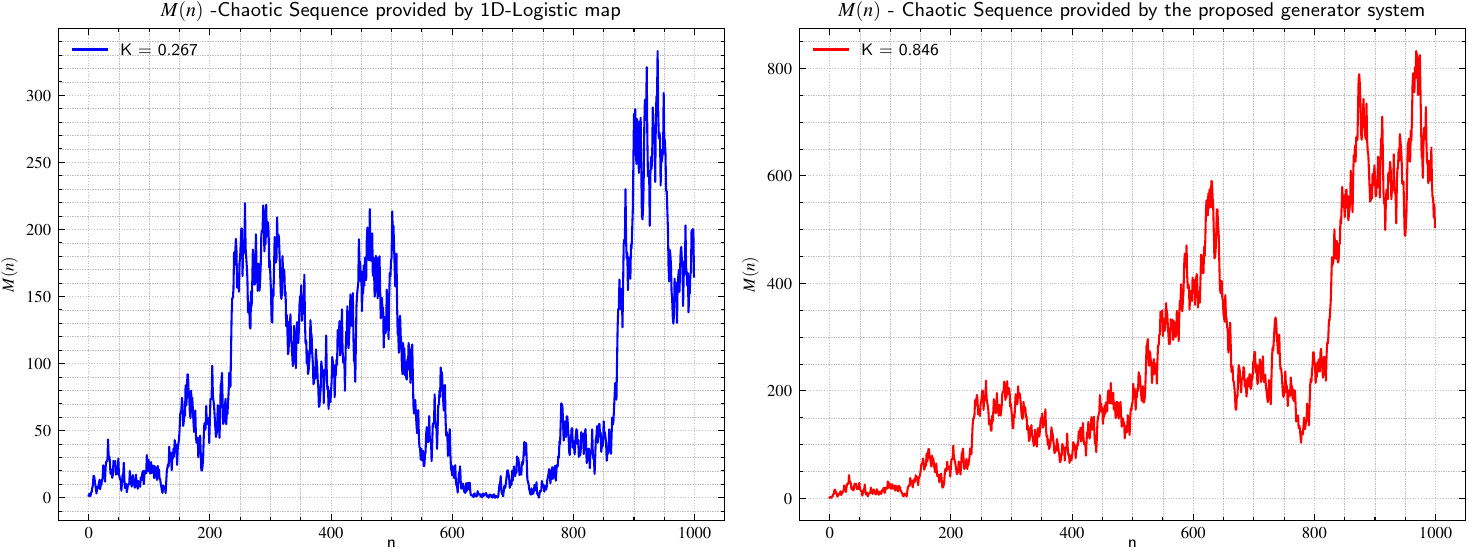}
	\caption{Mntrue and predicted}
	\label{Mn_true_vs_fully_predicted}
\end{figure}

\begin{figure}[ht]
	\centering
	\includegraphics[width=\linewidth]{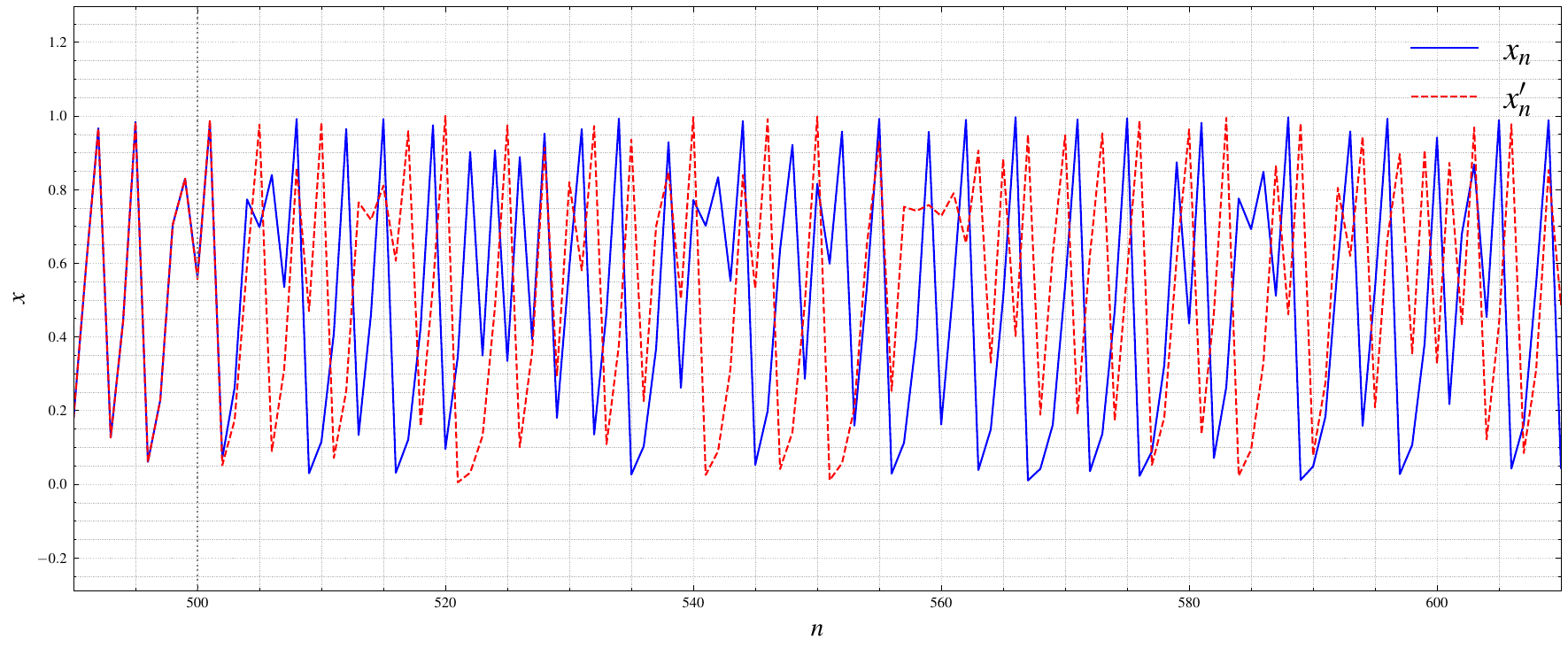}
	\caption{The difference of chaotic signals}
	\label{f_d}
\end{figure}

To further bolster the unpredictability of the generated sequences, we introduce the 2D Sin-Cos Pi Hyperchaotic Map (2D-SCPHM). As defined in \cref{q3}, this map is governed by parameters $a, b \in (0, 25)$ and distinctively incorporates sine and cosine functions, Gaussian mapping, and a mechanism for the dynamic control of parameters $a$ and $b$. A salient feature of the 2D-SCPHM is the integration of the transcendental number $\pi$ into its structure. This design choice aims to promote strong aperiodicity and enhance ergodicity, offering potential advantages over conventional chaotic maps such as the Logistic map \cite{qi2008new, leith1996stochastic}.

\begin{equation}
	\label{q3}
	\left\{
	\begin{aligned}
		x_{n+1} = a\sin(\pi^{y_{n}^2} - x_{n}^\pi);\\
		y_{n+1} = b\cos(\pi^{x_{n}^2} - y_{n}^\pi).\\
	\end{aligned}
	\right.
\end{equation}

The specific steps are as follows:

\textbf{Step 1: GRU model training.}
Before using GRU to generate sequences, we need to train the GRU model with the key. To begin, define the system parameters such as \( Key_{a} \) and set the initial value of the state variable \( Key_{x0} \). Using a fixed time step of 0.01, iteratively generate chaotic sequences from the specified chaotic system. To eliminate transient effects, discard the first 6,000 data points and retain the subsequent 24,000 points to form the pseudorandom sequence \( \{X\} \), which will be used for training and evaluation. Then, configure the structure and hyperparameters of the LSTM model—including the number of hidden units, learning rate, and sequence length—to enable effective time series learning. Finally, select a segment of the sequence \( \{X\} \) as the training data for the LSTM model. After obtaining the GRU model, we can generate a chaotic sequence.

\textbf{Step 1: Chaotic sequence generation.}
For a plaintext image \( P \) of dimensions \( M \times N \), the 2D-SCPHM (potentially influenced or initialized by the GRU component) is iterated \(\lceil (M \times N + N_0 + 3)/2 \rceil\) times. This iterative process employs the system parameters \( a, b \) and initial conditions \( (x_0, y_0) \). To mitigate transient effects inherent in chaotic systems, the initial \( N_0 \) pairs of values generated by the 2D-SCPHM are discarded. This yields two refined chaotic sequences, each of length \(\lceil (M \times N + 3)/2 \rceil\). The initial conditions \( x_0, y_0 \), system parameters \( a, b \), and the discard count \( N_0 \) constitute the secret keys for this module, denoted as $Key_{x0}$, $Key_{y0}$, $Key_{a}$, $Key_{b}$, and $Key_{N0}$, respectively.

\textbf{Step 2: Sequence partitioning.}
The two refined chaotic sequences, say \(S_x = \{x_i\}\) and \(S_y = \{y_i\}\), each of length \(\lceil (M \times N + 3)/2 \rceil\) as obtained from Step 1, are then interleaved or concatenated to form a composite sequence \(S_c\). This composite sequence \(S_c\) will have a total length of at least \( M \times N + 3 \). From this composite sequence \(S_c\), the initial \( M \times N + 3 \) values are selected. These are subsequently partitioned into two distinct sequences: (i) sequence \( X_s \), comprising the first \( M \times N \) values, designated for pixel permutation or diffusion operations; and (ii) sequence \( Y_{ctrl} \), consisting of the next 3 values, which are employed as dynamic control parameters in subsequent encryption phases.

\subsection{Knot-like Unique Novel-Scan Algorithm (Kun-SCAN)}

\begin{figure}[ht]
\centering
\includegraphics[width=\columnwidth]{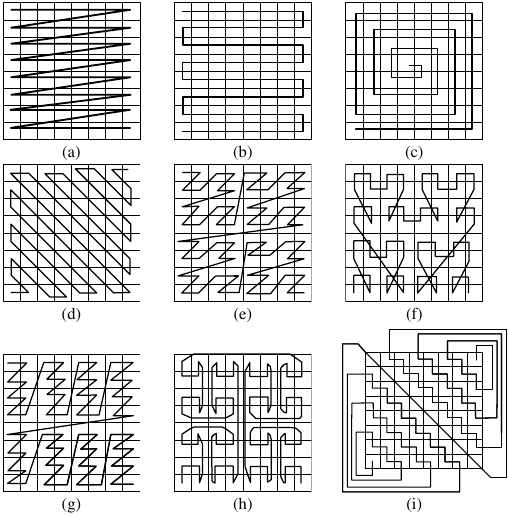}
\caption{Different Scan patterns: (a) Raster Scan, (b) Continuous Raster Scan, (c) Spiral Scan, (d) Diagonal Scan, (e) Z-order Pattern, (f) U-index Pattern, (g) Z-mirror Pattern, (h) Hilbert Curve, (i) Kun Scan}
\label{scans}
\end{figure}

Various image scanning methods, such as Raster, Spiral, and Hilbert curves (\cref{scans}), permute pixel positions to enhance confusion but often suffer from high local pixel correlation and vulnerability to statistical attacks. To overcome these limitations, we introduce the Knot-like Unique Novel-Scan Algorithm (Kun-SCAN). Kun-SCAN disrupts pixel continuity, reduces inter-pixel correlation, and eliminates periodicity through partitioned scanning, cross-boundary extraction, and multi-region traversal. It improves security against statistical and brute-force attacks and applies to various image types without prior segmentation.

\subsection{The XOR Diffusion Algorithm (XDA)}
Following the Kun-SCAN permutation stage, the scrambled image is transformed into a one-dimensional pixel sequence. This linearization is achieved by flattening the 2D pixel array, typically in a raster-scan order. This linearized pixel vector subsequently undergoes a nonlinear diffusion process, which is implemented using bitwise XOR operations. More precisely, each pixel value \(P'_i\) in the sequence is modified by XORing it with the preceding diffused pixel value \(C_{i-1}\) and a corresponding element \(K_i\) from a pseudorandom key stream (derived, for instance, from sequence \(X_s\) or another output of the chaotic generator). The diffusion can be expressed as \(C_i = P'_i \oplus C_{i-1} \oplus K_i\), where \(C_0\) (the initial value for the diffusion chain) can be a predetermined constant or derived from the secret keys. The iterative XOR diffusion strategy creates strong inter-pixel dependencies, ensuring that small changes in the plaintext propagate throughout the ciphertext. This enhances the algorithm's sensitivity to input changes, contributing to its overall security and aligning with the avalanche effect.

\subsection{The Decryption Algorithm}
The decryption procedure for Kun-IE is precisely the inverse of the encryption process. Each module's operations—XDA, Kun-SCAN, and chaotic sequence generation for key stream derivation—are reversed in sequence using the same set of secret keys ($Key_{x0}$, $Key_{y0}$, $Key_{a}$, $Key_{b}$, $Key_{N0}$). This ensures that the original plaintext images can be perfectly reconstructed from the ciphertext, provided the keys are identical and all operations are perfectly reversible.

\section{Experimental Results}
\label{Sec3}

This section presents a comprehensive evaluation of the proposed cryptosystem. The primary focus is on assessing the chaotic characteristics of the newly designed 2D Spatiotemporal Chaos-based Pseudo-Random Number Generator with Hyperchaotic Mapping (2D-SCPHM) and the overall security and efficiency of the Kun-IE encryption framework.

\subsection{Experimental Setup}

\begin{table*}[!ht]
    \centering
    \label{compare}
    \caption{Comparing coefficients of correlation schemes of one image.}
    \begin{tabular}{ccccccccc}
        \toprule
        \multirow{2}{*}{Algorithm} & \multicolumn{3}{c}{Correlation coefficients} & \multirow{2}{*}{Information entropy $\uparrow$} & \multirow{2}{*}{Time (s) $\downarrow$} & \multirow{2}{*}{NPCR} & \multirow{2}{*}{UACI} & \multirow{2}{*}{Key space $\uparrow$} \\ \cmidrule(lr){2-4}
        & Horizon $\downarrow$ & Vertical $\downarrow$ & Diagonal $\downarrow$ & & & & &  \\
        \midrule
        Proposed & \textbf{-0.0009} & \textbf{0.0011} & \textbf{0.0018} & \textbf{7.9993} & \textbf{0.3615} & \textbf{99.6097(*)} & \textbf{33.4639(*)} & \textbf{$2^{260}$} \\
	Lu \etal \cite{lu2020efficient} &0.0024&	0.0011& 0.0021& 7.9971 & 1.4890& 99.6554& 33.4665& $2^{124}$\\
	Kamal \etal \cite{kamal2021new}&	0.0145&	0.0115&	0.0087&7.9992&3.0901&	99.6010&33.4389&$2^{115}$\\
     Xu \etal \cite{xu2019fast}&	0.0067&	-0.0086&	0.0140&7.9935&0.5509&	99.6102&33.4812&$2^{30}$\\

        \bottomrule
    \multicolumn{9}{l}{ (*This value is closest to the ideal value)} \\
    \end{tabular}
\end{table*}

To evaluate the chaotic behavior of the 2D-SCPHM, it was compared with three contemporary 2D chaotic maps: 2D-TM, 2D-SSCDB, and Cross-2DHM (\cref{comparechaotic}). The evaluation used two medical image datasets—BCSS and Brain Tumor MRI—testing Kun-IE on six images and comparing its performance with established methods using standard security and performance metrics. All experiments were run on a workstation with an AMD Ryzen 7 7800X3D, 32 GB RAM, and Matlab R2024a.

\subsection{Performance Analysis of the 2D-SCPHM Chaotic Map}
This subsection details the experimental evaluation of the proposed 2D-SCPHM, focusing on its dynamical properties critical for cryptographic applications.

\begin{table}[ht]
    \caption{Correlation comparison of different scanning methods in four directions.}
    \centering
    \label{tab:scanning_comparison}
    \adjustbox{width=\columnwidth}{
    \begin{tabular}{ccccc}
        \toprule \text{Scanning Method} & Horizontal $\downarrow$ & Vertica l$\downarrow$ & Diagonal $\downarrow$ & Anti-Diagonal $\downarrow$\\
        \midrule \text{Raster Scan} & 0.9908 & 0.6830 & 0.6840 & 0.6651 \\
         \text{Continuous Raster Scan} & 0.9884 & 0.9910 & 0.9792 & 0.9769 \\
         \text{Spiral Scan} & 0.9920 & 0.4550 & 0.4577 & 0.4149 \\
         \text{Zigzag Scan} & 0.9801 & 0.2882 & 0.2631 & 0.2821 \\
         \text{Z-order Scan} & 0.9809 & 0.3418 & 0.3570 & 0.3525 \\
         \text{Z-mirror Scan} & 0.9884 & 0.7156 & 0.7222 & 0.7188 \\
         \text{Gray Scan} & 0.9803 & 0.4854 & 0.5020 & 0.5151 \\
         \text{Hilbert Scan} & 1.0000 & 0.9982 & 0.9985 & 0.9981 \\
         \text{U-index Scan} & 0.9890 & 0.9132 & 0.8791 & 0.9123 \\
         \text{Proposed Kun-SCAN} & \textbf{0.9625}& \textbf{0.0226} & \textbf{0.0077} & \textbf{0.0016} \\
        \bottomrule
    \end{tabular}}
\end{table}

\begin{table}[ht]
	\centering
	\caption{Some 2D hyperchaotic maps that used in the comparison.}
    \label{comparechaotic}
    \adjustbox{width=\columnwidth}{
	\begin{tabular}{llcc}
		\toprule
		Name  & F($x$, $y$) & Parameters & Ref  \\ \midrule
		2D-TM& \( \left\{ \begin{aligned}
			x_{n+1}&=\sin\left(\omega x_n\right)-r \sin \left(\omega y_n\right) \\
			y_{n+1}&=\cos\left(\omega x_n\right)
		\end{aligned} \right. \) & $\omega, r$ & \cite{tsafack2020new}\\
		2D-SSCDB& \( \left\{ \begin{aligned}
			x_{n+1}&=\sin \left(\mu x_n\left(1-y_n\right)+1\right), \\
			y_{n+1}&=\sin \left(\frac{\eta}{\left(x_n+y_n\right)}+1\right)\\  
		\end{aligned} \right. \) &$\mu, \eta$ &\cite{hu2023novel} \\ 
		\begin{tabular}[t]{@{}l@{}}
			Cross-2DHM \\ 
		\end{tabular} & \( \left\{ \begin{aligned}
			x_{i+1}&=\sin \left(\frac{\alpha}{\sin \left(y_i\right)}\right) \\
			y_{i+1}&=\beta \sin \left(\pi\left(x_i+y_i\right)\right)
		\end{aligned} \right. \) & $\alpha, \beta$ & \cite{teng2021color}\\
		
		2D-SCPHM& \( \left\{ \begin{aligned}
			%
			x_{n+1} = a\sin(\pi^{y_{n}^2} - x_{n}^\pi);\\
		y_{n+1} = b\cos(\pi^{x_{n}^2} - y_{n}^\pi);\\
		\end{aligned} \right. \) &$a, b$ & - \\
		\bottomrule
	\end{tabular}}
	\label{t111}
\end{table}

\subsubsection{Bifurcation Analysis}
\label{s1}
\begin{figure}[!h]
    
    \centering
    \subfigure[]   
    {\begin{minipage}[b]{.3\linewidth}        
            \centering
            \includegraphics[scale=0.2]{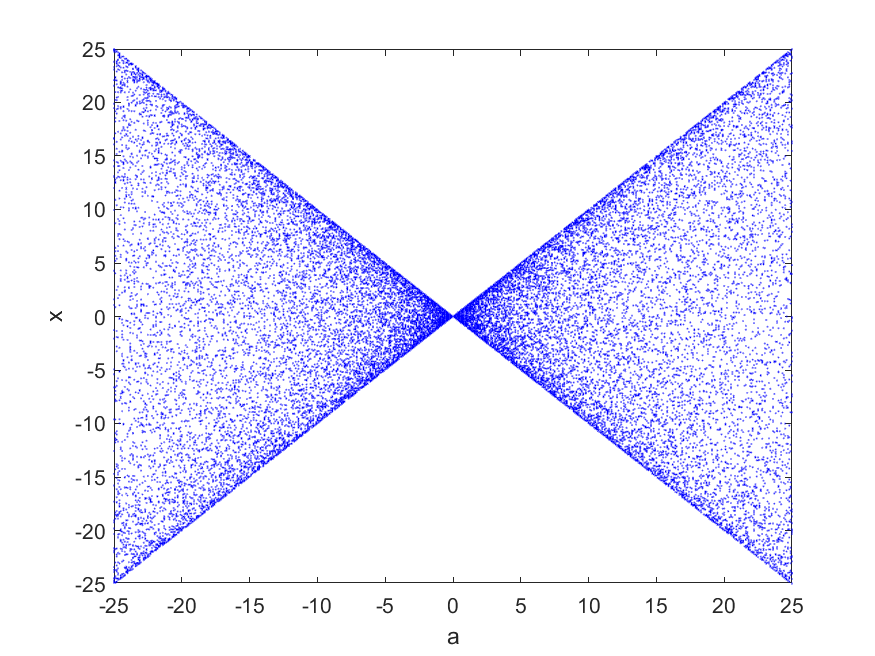}
            \label{hypergaussincubicbifax}
        \end{minipage}  
    } 
    \subfigure[] 
    {\begin{minipage}[b]{.3\linewidth}        
            \centering
            \includegraphics[scale=0.2]{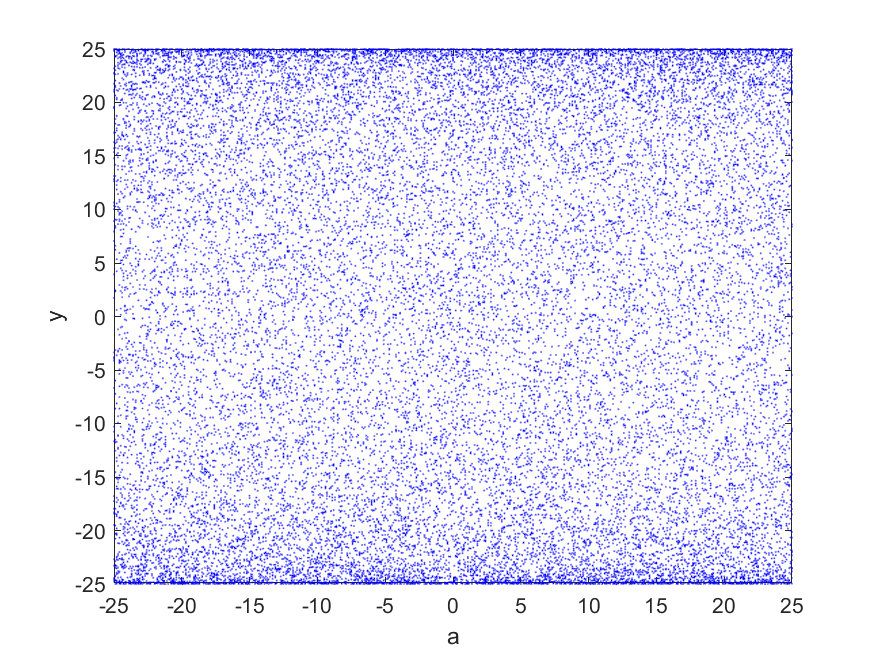}
            \label{hypergaussincubicbifay}
        \end{minipage} 
    }
    \subfigure[]
    {\begin{minipage}[b]{.3\linewidth}
            \centering
            \includegraphics[scale=0.2]{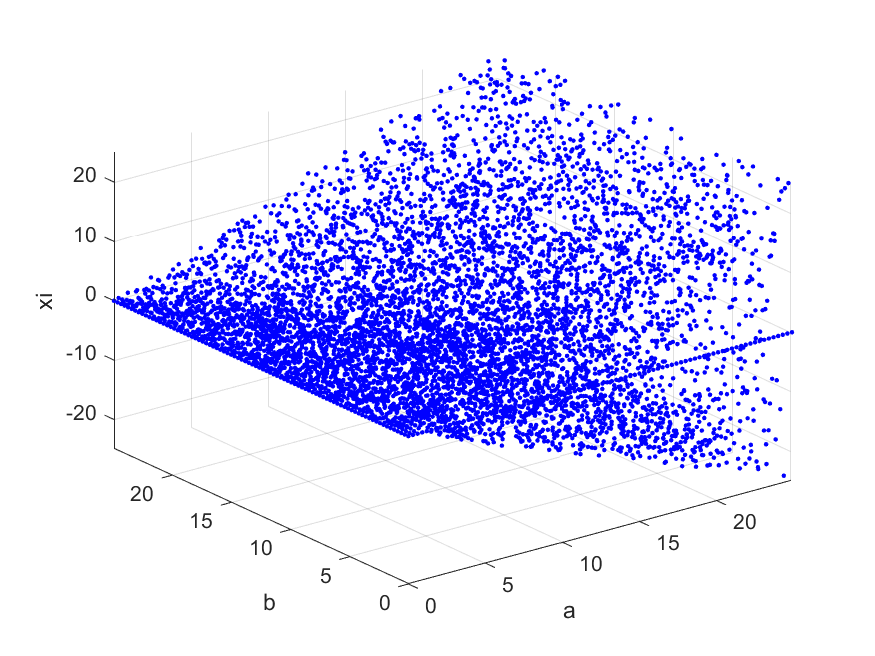}
            \label{hypergaussincubicbif1}
        \end{minipage}   
    }
    \subfigure[]   
    {\begin{minipage}[b]{.3\linewidth}        
            \centering
            \includegraphics[scale=0.2]{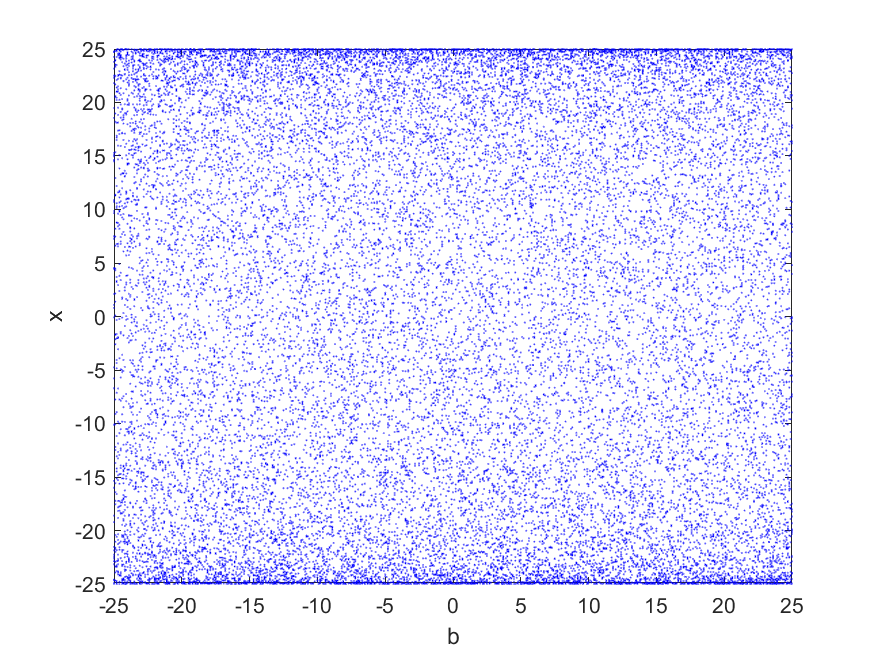}
            \label{hypergaussincubicbifbx}
        \end{minipage}  
    } 
    \subfigure[] 
    {\begin{minipage}[b]{.3\linewidth}        
            \centering
            \includegraphics[scale=0.2]{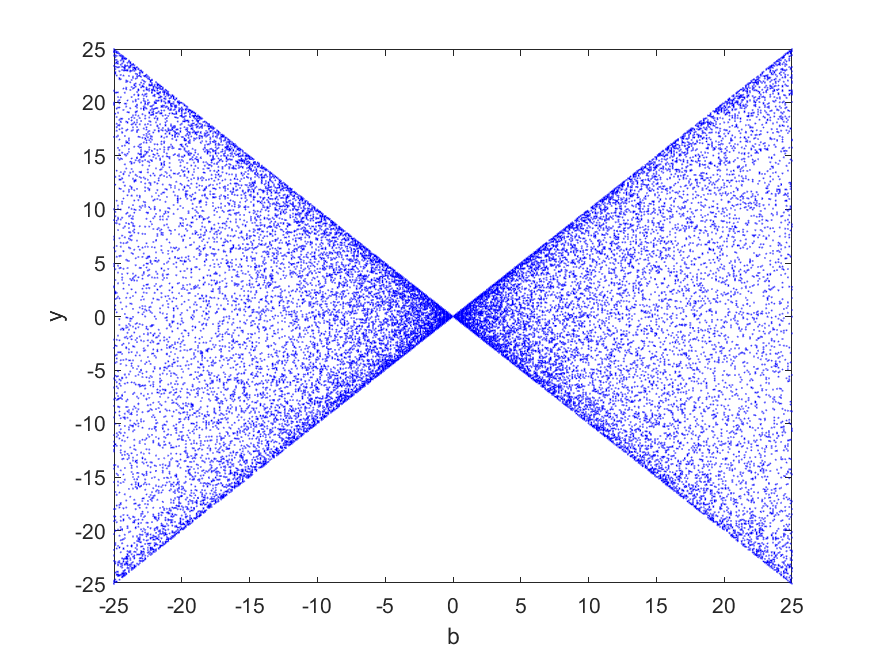}
            \label{hypergaussincubicbifby}
        \end{minipage} 
    }
    \subfigure[]
    {\begin{minipage}[b]{.3\linewidth}
            \centering
            \includegraphics[scale=0.2]{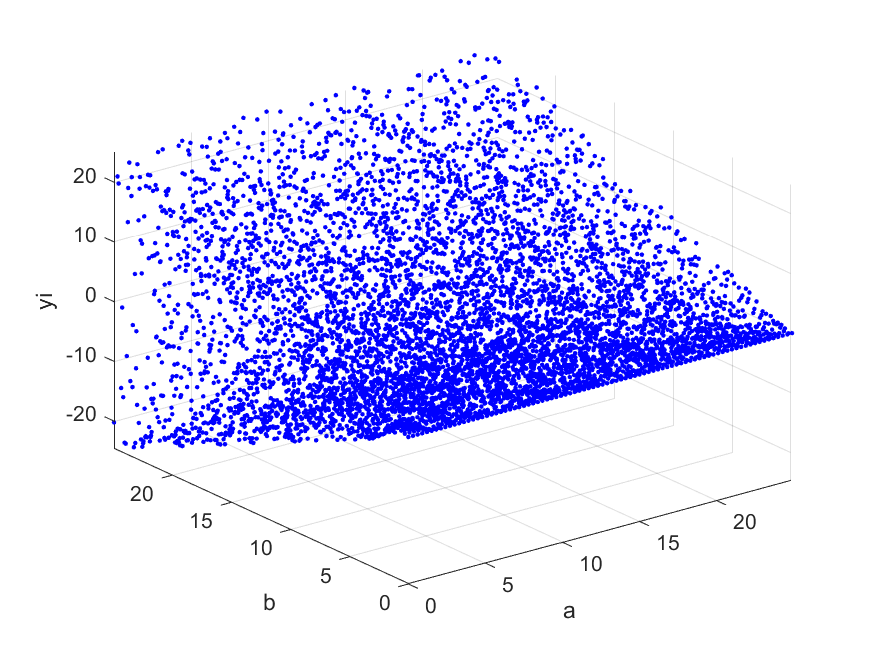}
            \label{hypergaussincubicbif2}
        \end{minipage}   
    }
    \caption{Bifurcation diagrams of the 2D-SCPHM: (a) and (b) depict 2D diagrams with $b$ = 25; (d) and (e) depict 2D diagrams with $a$ = 25; (c) and (f) depict 3D diagrams.}
    \label{Bifurcation}
    
\end{figure}

Bifurcation diagrams and attractor trajectories are instrumental in visualizing the dynamical characteristics of chaotic systems. As depicted in \cref{Bifurcation}, the 2D-SCPHM exhibits four distinct bifurcation diagrams in two-dimensional projections and two in three-dimensional representations. These diagrams collectively indicate a consistently uniform distribution of state values across the phase space, a desirable feature for chaotic maps used in cryptography.

\subsubsection{Lyapunov Exponent Analysis}
\begin{figure}[!h]
	\centering
	{
		\subfigure[2D-TM]
		{\begin{minipage}[b]{.2\linewidth}
				\centering
				\includegraphics[width=\linewidth]{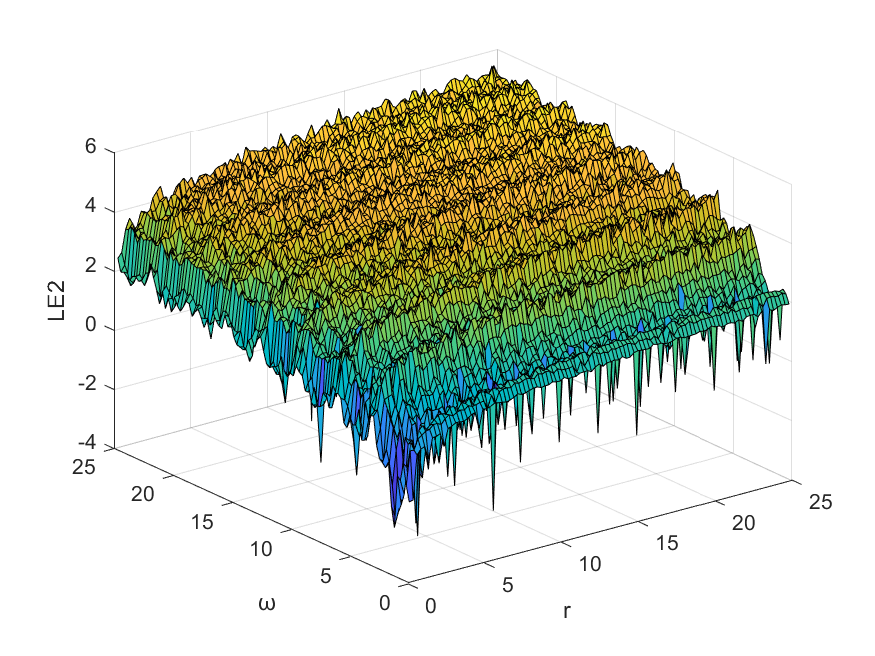}
				\includegraphics[width=\linewidth]{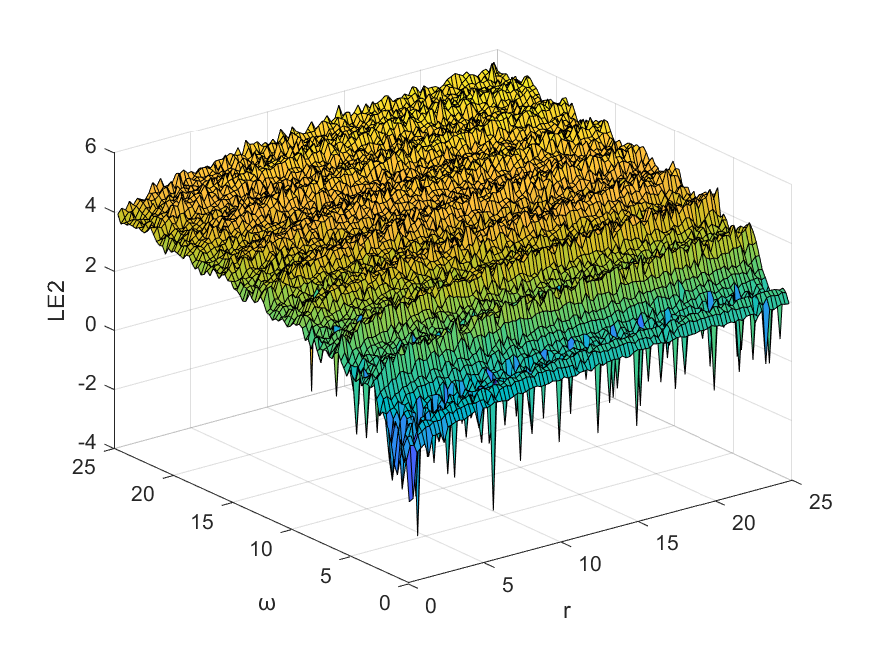}
			\end{minipage}   
		}
		\subfigure[2D-SSCDB]
		{\begin{minipage}[b]{.2\linewidth}
				\centering
				\includegraphics[width=\linewidth]{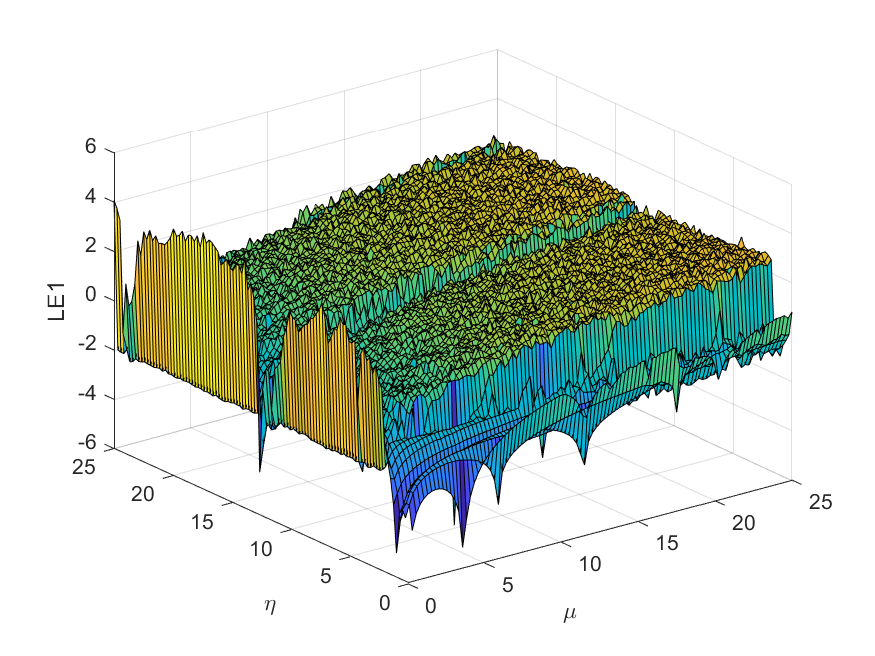}
				\includegraphics[width=\linewidth]{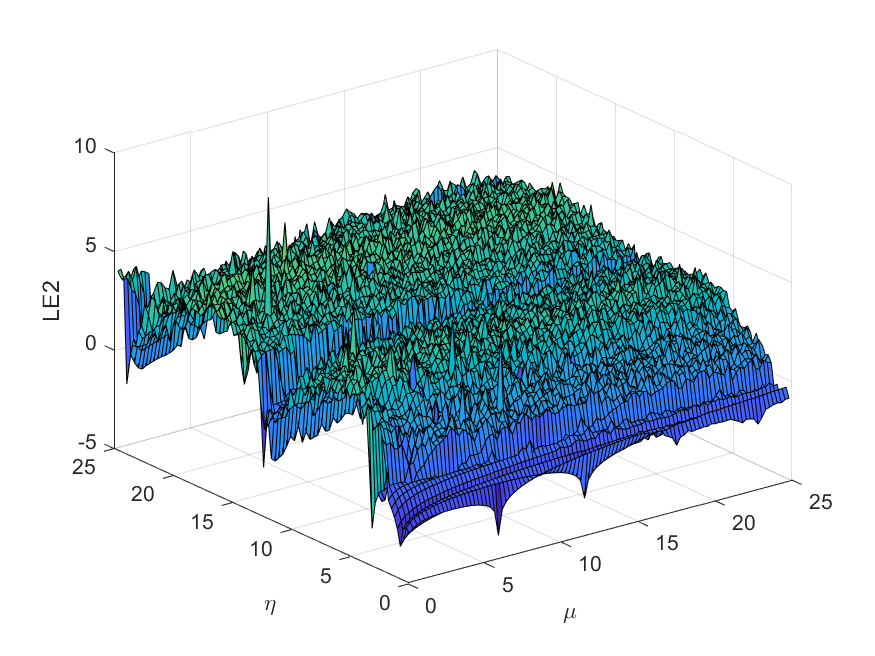}
			\end{minipage}   
		}
		\subfigure[Cross-2DHM]
		{\begin{minipage}[b]{.2\linewidth}  
				\centering
				\includegraphics[width=\linewidth]{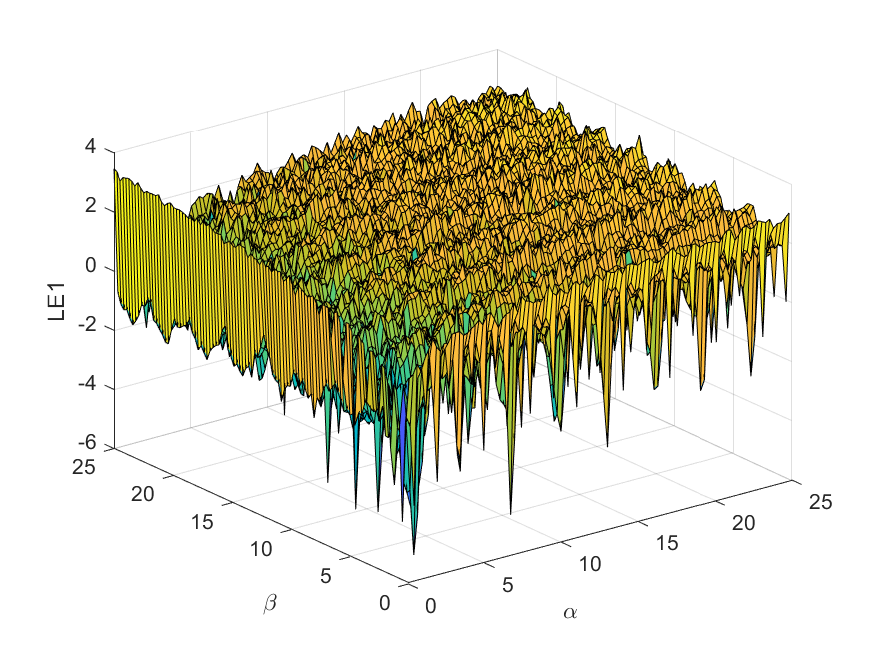}
				\includegraphics[width=\linewidth]{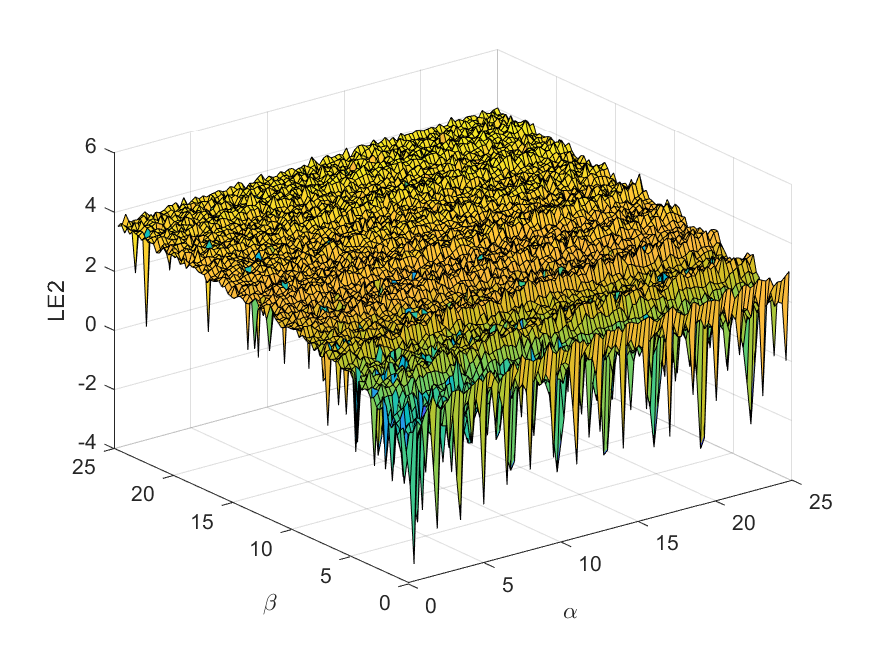}
			\end{minipage}   
		}
		\subfigure[2D-SCPHM]
		{\begin{minipage}[b]{.2\linewidth}
				\centering
				\includegraphics[width=\linewidth]{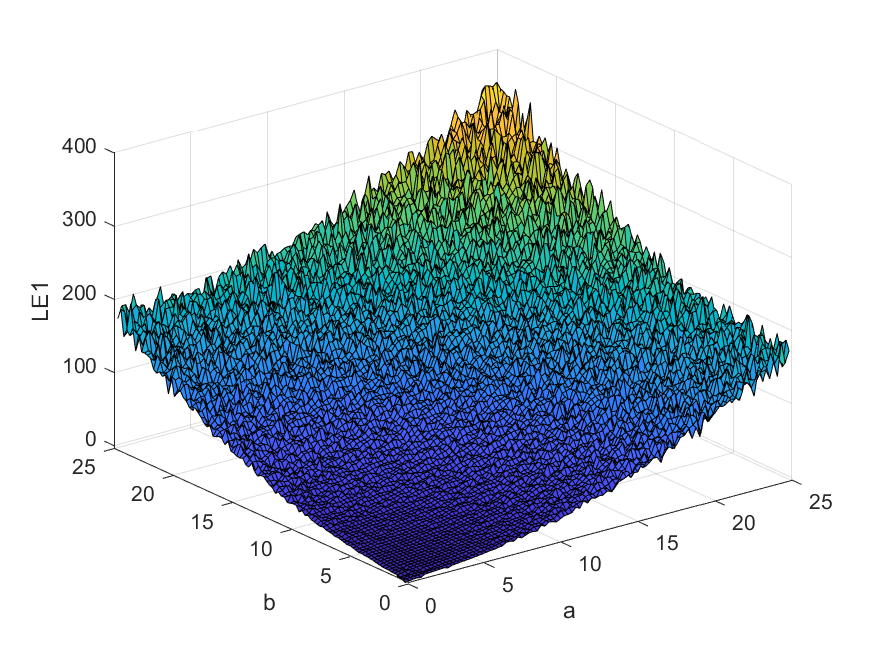}
				\includegraphics[width=\linewidth]{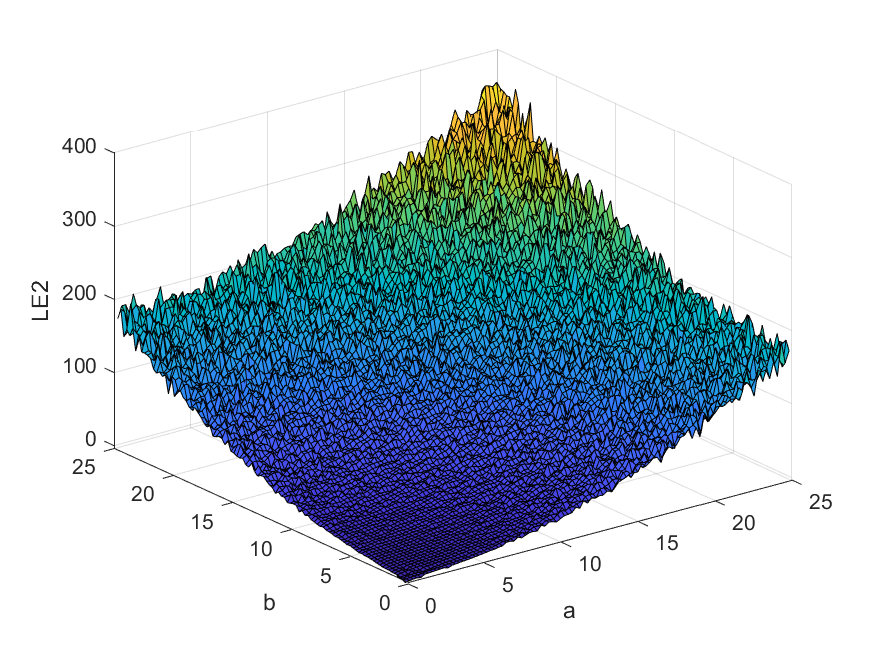}
			\end{minipage}   
		}
	}
	\caption{The three-dimensional Lyapunov exponent for various 2D chaotic maps.}
	\label{SELy}
\end{figure}

The Lyapunov exponent (LE) is crucial for identifying chaotic behavior. \Cref{SELy} compares the LEs of 2D-SCPHM and other maps in a 3D plot. The analysis varied $a$ and $b$ in the range (0, 25) to observe LEs (LE1 and LE2), with average values listed in \cref{comparehypermap}. Notably, the maximum LE for 2D-SCPHM reaches nearly 400, surpassing that of 2D-TM, 2D-SSCDB, and Cross-2DHM, highlighting its strong chaotic properties and suitability for chaotic sequence generation.

\begin{table}[ht!]   
	\caption{Average values of LE1 and LE2 for 2D-SCPHM and other chaotic maps.}
	\centering
	\begin{tabular}{cccc}
		\toprule
		\multirow{2}{*}{Map name} & \multirow{2}{*}{Control Parameters} & \multirow{2}{*}{LE1} & \multirow{2}{*}{LE2} \\
		& & & \\ \midrule
		2D-TM       & $\omega, r$       & 5.0573    & 1.9086 \\
		2D-SSCDB    & $\mu, \eta$       & 3.2324    & 1.4238 \\
		Cross-2DHM  & $\alpha, \beta$   & 3.1922    & 3.6422 \\
		2D-SCPHM    & $a, b$            & 233.5333  & 236.4123 \\
		\bottomrule
	\end{tabular}
	\label{comparehypermap}
\end{table}

\subsubsection{NIST SP 800-22 Randomness Tests}
\label{s2}
The NIST SP 800-22 suite includes fifteen tests evaluating the randomness of binary sequences, where $P$-values must exceed 0.01. As shown in \cref{nist}, sequences generated by 2D-SCPHM passed all tests, confirming its suitability for cryptographic use and high-quality pseudo-random number generation.

\begin{table}[ht]
        \centering
		\caption{NIST test results of 2D-SCPHM.}
        \adjustbox{width=\columnwidth}{
		\begin{tabular}{llccc}
			\toprule
			Test index & $P$$-$value of $x$ sequence & $P$$-$value of $y$ sequence & Pass &  \\ \midrule
			\multirow{1}{*}{Frequency}    & 0.2622				& 0.0909		&2/2    &  \\ 
			\multirow{1}{*}{BlockFrequency}   &0.7399     		& 0.0757		&2/2    &  \\ 
			\multirow{1}{*}{Cumulative Sums}    &0.2905			& 0.4950		&2/2    &  \\ 
			\multirow{1}{*}{Runs}       & 0.0077      			& 0.3345		&2/2    &  \\ 
			\multirow{1}{*}{Longest Run}       &0.2757    		& 0.2757		&2/2    &  \\ 
			\multirow{1}{*}{Binary Matrix Rank}    & 0.6163		& 0.0457		&2/2    &  \\ 
			\multirow{1}{*}{Discrete Fourier Transform}    &0.7792 & 0.1373		&2/2  	&  \\ 
			\multirow{1}{*}{Non-overlapping Template}    &0.5048 	& 0.4853		&2/2    &  \\ 
			\multirow{1}{*}{Overlapping Template}    &0.3041 	& 0.2133		&2/2    &  \\ 
			\multirow{1}{*}{Maurers Universal Statistical}    &0.6579		& 0.3375		&2/2    &  \\ 
			\multirow{1}{*}{Approximate Entropy}    &0.3669 	& 0.3231		&2/2    &  \\ 
			\multirow{1}{*}{Random Excursions}    &0.4113	& 0.4750		&2/2    &  \\ 
			\multirow{1}{*}{Random Excursions Variant}    &0.3142 	& 0.0628		&2/2    &  \\ 	
			\multirow{1}{*}{Serial Test}    &0.5803 	& 0.8959		&2/2    &  \\ 
			\multirow{1}{*}{Linear Complexity Test}    &0.5544		& 0.2757		&2/2   	&  \\ 
			\bottomrule	
		\end{tabular}}
		\label{nist}
\end{table} 

\subsubsection{Generation of Novel Chaotic Sequences using GRU}
To explore the potential for generating extended and novel chaotic sequences, a Gated Recurrent Unit (GRU) network was trained using time series data produced by 2D-SCPHM. The training performance, as shown in \cref{Mn_true_vs_fully_predicted}, indicates that the GRU model achieves a low root-mean-square error (RMSE) and loss function value, signifying its capability to effectively learn the underlying dynamics of the chaotic system. \Cref{f_d} demonstrates that the sequences $\{X'\}$ generated by the trained GRU, while preserving the essential chaotic characteristics of the original 2D-SCPHM output, are novel and distinct. This capability is valuable for applications requiring a large volume of unpredictable sequences, such as one-time pads or diverse key stream generation. (Note: The original text mentioned LSTM in the body but GRU in the title; this version consistently uses GRU. Please verify and adjust if LSTM was intended.)

\subsection{Security and Performance Analysis of Kun-IE}
This subsection evaluates the proposed Kun-IE cryptosystem in terms of its security robustness and computational efficiency.

\subsubsection{Key Space Analysis}
A secure cryptosystem requires a sufficiently large key space to prevent brute-force attacks. Traditionally, a key space larger than $2^{100}$ is considered secure. Kun-IE uses five secret keys with ranges for initial conditions, control parameters, and system parameters, resulting in an effective key space of $2^{260}$. As shown in \Cref{compare}, Kun-IE is highly resistant to brute-force attacks, ensuring strong security.

\begin{figure}[ht]
	\centering
\subfigure[]
{\includegraphics[width=0.32\linewidth]{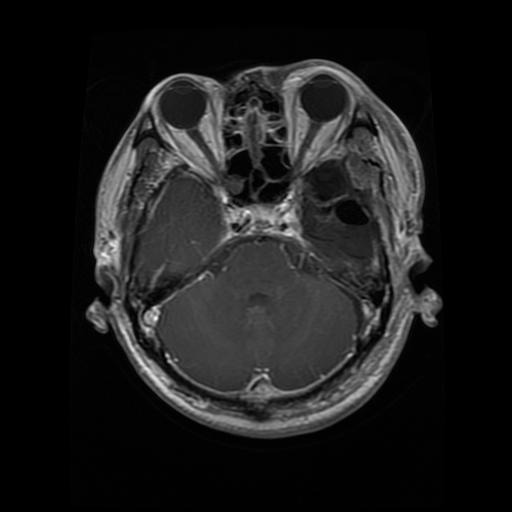}}
\subfigure[]
{\includegraphics[width=0.32\linewidth]{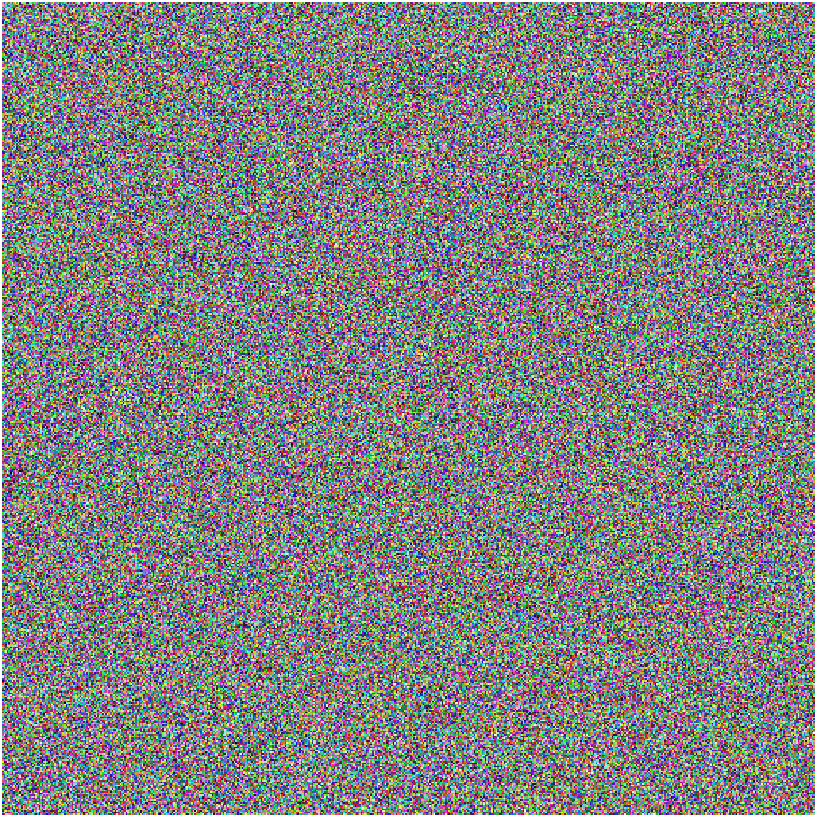}}
\subfigure[]
{\includegraphics[width=0.32\linewidth]{Fig/1002.jpg}}
\subfigure[]
{\includegraphics[width=0.32\linewidth]{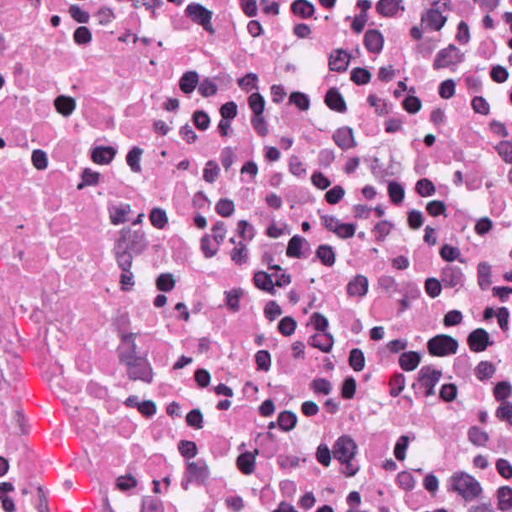}} 
\subfigure[]
{\includegraphics[width=0.32\linewidth]{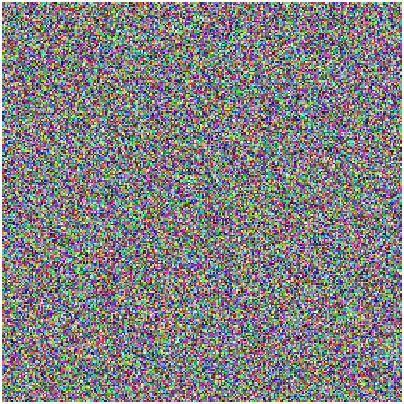}}
\subfigure[]
{\includegraphics[width=0.32\linewidth]{Fig/TCGA-A1-A0SK-DX1_xmin45749_ymin25055_MPP-0_1536_1024_size512.png}} 
\caption{Simulation results: (a,d) The original images of Image1, Image2, and Image3 respectively; (b,c) The encrypted images of (a,d), respectively; (c,f) The decrypted images of (b,c).}
\label{figure_histograms}
\end{figure}

\subsubsection{Sensitivity Analysis}
Differential cryptanalysis examines how slight key changes affect ciphertext. Sensitivity, measured by NPCR and UACI, reflects resistance to differential attacks. As shown in \Cref{compare}, the proposed method shows high key sensitivity; its NPCR ($\approx$ 99.6\%) and UACI ($\approx$ 33.4\%) closely match ideal values, confirming strong differential security.

\subsubsection{Information Entropy Analysis}
Information entropy $H(m)$ measures the randomness of an image. For an 8-bit grayscale image, the theoretical maximum is 8 bits/pixel. As shown in \cref{compare}, Kun-IE’s encrypted images achieve entropy values near 8, indicating high randomness and effective concealment of plaintext information.

\subsubsection{Correlation Coefficient Analysis}
Natural images show strong correlation among adjacent pixels. A secure encryption scheme must suppress these correlations to resist statistical attacks. As shown in \Cref{compare}, plaintext images have correlation coefficients near 1, while Kun-IE ciphertext values are close to 0, demonstrating its effectiveness in eliminating pixel correlation and concealing image statistics.

\subsubsection{Performance Analysis of the Kun-SCAN Algorithm}
The scrambling performance of Kun-SCAN was compared with other scanning methods using the Lena image ($512 \times 512$ grayscale). Each image was converted to a 1D sequence and reconstructed for invertibility, with the process repeated three times. Correlation coefficients in four directions were then computed, showing Kun-SCAN consistently produced the lowest inter-pixel correlation, confirming its superior scrambling ability and contribution to encryption security.

\subsubsection{Encryption Time Evaluation}
A practical cryptosystem must be both secure and efficient. Kun-IE’s encryption speed was benchmarked against other algorithms under identical conditions. As shown in \cref{compare}, Kun-IE encrypts images rapidly, confirming its suitability for real-time and practical applications.

\begin{table*}[H]
	\centering
	\caption{Encryption time (seconds).}
    	\label{t12}
	\begin{tabular}{cccccccc}
		\toprule
		{Image size} 
		&$128 \times 128 \times 3$ &$256 \times 256 \times 3$ &$512 \times 512 \times 3$ &$1024 \times 1024 \times 3$ \\ \midrule
		{Encryption time}
		& 0.0197 & 0.0816&0.3615  &1.6098
		\\ \bottomrule
	
	\end{tabular}

\end{table*}

\section{Conclusion}
\label{Sec4}

To improve key and plaintext sensitivity in encryption schemes, we propose a new chaotic system (2D-SCPHM) and a diffusion mechanism (JNTM) for color image encryption. 2D-SCPHM generates pseudo-random sequences, while the extended Arnold Transform scrambles the image. JNTM applies forward and backward diffusion to ensure that any plaintext change affects the whole image. Experiments demonstrate the scheme's strong resistance to brute-force, statistical, differential, noise, and chosen/known-plaintext attacks, confirming its high security and efficiency in image encryption.

\bibliographystyle{IEEEtran} 
\bibliography{mycite}   

\end{document}